\documentclass[twocolumn,showpacs,preprintnumbers,amsmath,amssymb]{revtex4}
\usepackage{mathrsfs}
\usepackage{dcolumn}
\usepackage{bm}
\usepackage{graphicx}
\begin{document}
\title{Phase-controlled localization and directed transport in a bipartite lattice}
\author{Kuo Hai$^{1}$\footnote{ron.khai@gmail.com},\ \ Yunrong Luo$^{1}$,\ \ Gengbiao Lu$^{2}$,\ \ Wenhua Hai$^{1}$\footnote{
whhai2005@yahoo.com.cn}}
\affiliation{$^{1}$Department of physics and Key Laboratory of
Low-dimensional Quantum Structures and \\
Quantum Control of Ministry of Education,
Hunan Normal University, Changsha 410081, China
\\ $^{2}$Department of Physics and Electronic Science, Changsha University of Science and Technology,
Changsha 410004, China}

\begin{abstract}

We investigate coherent control of single particles held in a bipartite optical lattice via a combined high-frequency modulation. Our analytical results show that for the photon resonance case the quantum tunneling and dynamical localization depend on the phase difference between the modulation components, which leads to a different route of the coherent destruction of tunneling and a simple method for stabilizing the system to implement the directed transport. The results could be referable for manipulating the transport characterization of the
similar tilted and shaken optical or solid-state systems, and also can be extended to the
many-particle systems.


\end{abstract}

\maketitle

\section{Introduction}

Quantum control of tunneling processes of single particles plays a major role in different areas of physics and chemistry \cite{Rabitz, Grifoni, kral} ranging
from Shapiro steps in Josephson junctions \cite{Shapiro} to the control of chemical reactions via light in molecules \cite{Tannor}. As early as 1986, Dunlap and Kenkre studied theoretically
the quantum motion of a charged particle on a discrete lattice driven by an ac field \cite{Dunlap}, and found the surprising
result that particle transport can be completely
suppressed when ratio of the strength and the frequency of the ac field takes some special
values. This effect of dynamical localization (DL) was
later found to be associated with the coherent destruction of tunneling (CDT) \cite{Grossmann, Grifoni} at a collapse point of the Floquet quasienergy spectrum \cite{Holthaus}, and has also been observed in different systems \cite{Madison, Lignier, Valle, E.Kierig}.
There has been growing interest in the quantum
control of electrons in semiconductor superlattices or arrays of
coupled quantum dots from both theoretical and experimental sides
\cite{Gluck, Grifoni, Keay, Villas,Saito}. Most of the DL and CDT of the electronic
systems are generic and also can occur in atomic
\cite{Madison, Holthaus, Lignier, Weiss} and optical \cite{Valle, Longhi2,Dreisow,Xie}
systems.

Recently, different routes to CDT were found by considering, respectively, the priori prescribed number of bosons of a many-boson system \cite{Gong, Longhi1}, the distinguishable intersite separations of a bipartite lattice \cite{Creffield2,Kuo}, the variable driving symmetry of a two-frequency driven particle in a double-well \cite{E.Kierig, LuG}, and the different combined modulations to a two-level system \cite{Hai}. The CDT mechanism has been applied to different physical fields such as the quantum transition in ultracold atomic systems \cite{Zenesini,Eckardt}, the directed transport in a bipartite lattice \cite{Creffield2,Kuo}, and the design of quantum tunneling switch based on a planar four-well \cite{Lu2}.
It is worth noting that the CDT mechanism can also be applied to coherently control instability of a periodically driven
system \cite{Xiao}. Stability of a quantum state has been recognized long ago during the heyday of quantum mechanics. In the sense of Lyapunov, by the instability of a solution we mean that the initially small deviations from the given solution
grow without upper limit that could lead to destruction of the
solution behavior. Therefore, investigation on the stability is important for the practical application.
The instability of the periodically driven
lattice systems has been investigated \cite{Creffield,khai}. It is found that stability of the systems depends on signs of the effective tunneling rates of two nearest-neighbor barriers such that one can stabilizes the systems by tuning the effective tunneling rates. Here our aim is finding a new route of CDT and supplying a simple stabilization method for transporting single particles in a driven bipartite lattice.

The coherent control of an ac driven particle in a lattice
with a single intersite separation has been investigated widely in the
nearest-neighbor tight binding (NNTB) approximation \cite{Dunlap,Rivera,Longhi}. More recently, a bipartite lattice or double-well train with two different intersite separations \cite{Qian} is applied to induce the ratchetlike effect \cite{Creffield2,Kuo}, to transport quantum
information \cite{Romero} and to realize two-qubit quantum gates \cite{Chiara}.
The periodic modulation is usually applied to
the potential tilt (bias) between the lattice sites \cite{Dunlap,Rivera,Longhi} or the tunnel coupling \cite{Massel, RMa, Chen}. For an analytically solvable two-site system, combined modulations have been applied to produce the exact solutions \cite{Hai,Hioe}. The periodic modulation can be performed in a nonadiabatic \cite{Creffield2,Romero} or adiabatic manner \cite{Qian,SLonghi}.

In this work, we consider single particles held in an optical bipartite
lattice with two different separations $a$ and $b$ and driven by a combined modulation of two resonant
external fields with a phase difference between the bias and coupling. In the high-frequency regime and NNTB approximation, we derive an
analytical general solution for the probability amplitude of the particle
in any localized state in which the characterization of quantum tunneling and stability depend on the phase difference between the modulation components. A new route of CDT and a simple method for stabilizing the system to perform the directed transport are found by adjusting the phase difference adiabatically \cite{Qian,SLonghi} or nonadiabatically \cite{Creffield2,Romero}. Such a phase-adjustment may be more convenient in experiments compared to the usual amplitude- and frequency-modulations. The results can be tested with existing experimental setups on the periodically tilted and shaken optical lattices and could be applied to simulating the similar optical systems \cite{Longhi2,Dreisow} and solid-state systems \cite{Villas,Longhi}. At the end of the paper, we suggest a scheme for extending the results to a
many-particle system.

\section{General solution in the high-frequency regime}

We consider a driven and tilted bipartite lattice (double-well train) of form $V(x,t)=V_1(t)\cos (k_L x)+V_2(t)\cos(2 k_L x)+g(t) x$ with time-periodic lattice depths \cite{Chen,Romero} $V_j(t)$ and potential tilt \cite{Madison,Zenesini} $g(t)$ between the lattice sites, which consists of the tilted long lattice of wave-vector $k_L$ and short lattice of wave-vector $2k_L$. Such a lattice
can be realized experimentally by a periodically shaken optical lattice \cite{Chen}, and by imposing a phase modulation to one of the standing wave component fields \cite{Madison} or by moving the position of a retroreflecting mirror
which is mounted on a piezoelectric actuator \cite{Zenesini}. A single particle is initially placed near the lattice center, as shown in Fig. 1, where the different separations $a$ and $b$ are adjusted by the laser wave vector $k_L$ and amplitudes. Here we have selected a suitable initial time $t_0$ and phase difference $\phi$ between $V_j(t)$ and $g(t)$ to make $g(t_0)=0$ and $V_j(t_0)\ne 0$. Quantum dynamics of such a
system is governed by the Hamiltonian \cite{Creffield2,khai}
\begin{eqnarray}
H(t)=\sum_{(i,j)}J(t)(b_i^\dag b_j+H.C.)- \mathscr{E}(t)\sum_n
x_nb_n^+b_n.
\end{eqnarray}
Here $(i, j)$ means the nearest-neighbor site pairs, $J(t)=J_0+\delta J \cos(m \omega t-\phi)$ for $m=0,1,2,...$
denotes the tunnel coupling \cite{Massel, RMa, Chen} and $\mathscr{E}(t)=\mathscr{E}_0 \cos(\omega t)$  is the potential tilt \cite{Dunlap, Madison}, where $J_0$ is a constant, $\delta J,\ \mathscr{E}_0$ and $\omega$ are the driving intensities and frequency, $m\ne 0$ means the photon resonance. Signs $b^\dag_j$ and $b_j$ are,
respectively, the particle creation and annihilation operators in
the site $j$. The spatial locations of the $n$th lattice
sites read $x_n=n(a+b)/2$ for
even integer $n$, and $x_n=(n+1)a/2+(n-1)b/2$ for odd $n$. To
simplify, we have set $\hbar=1$ and normalized energy and time
by $\omega_0$ and $\omega_0^{-1}$ with $\omega_0$ being a fixed
reference frequency in order of $J_0$. The parameters $J_0,\ \delta J$ and
$(\mathscr{E}_0x_n)$ are in units of $\omega_0$ with $x_n$ being
normalized by the fixed reference length $l_0\sim 1\mu$m. Thus all
the parameters are dimensionless throughout this paper.

\begin{figure}
\begin{center}
\includegraphics[width=2.5in]{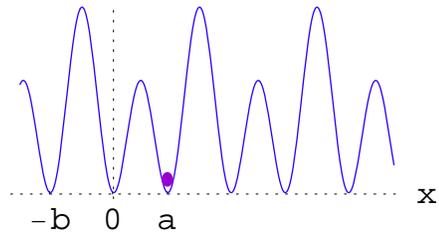}
\caption{A single particle is initially placed in the driven bipartite lattice
centered at coordinate $0$ with two different separations $a$ and
$b$. Hereafter all the quantities plotted in the figures are
dimensionless.}
\end{center}
\end{figure}

Letting $|n \rangle$ be the localized state at the site $n$, we
expand the quantum state $|\psi(t)\rangle$ as the linear
superposition $|\psi(t)\rangle=\sum_n c_n(t)|n\rangle$. Combining
this with Eq. (1), from the time-dependent Schr\"{o}dinger equation
$i\frac{\partial}{\partial t}|\psi(t)\rangle=H(t)|\psi(t)\rangle$ we
derive the coupled equations of the probability amplitudes \cite{khai,
SLonghi}
\begin{eqnarray}
i\dot{c}_n(t)=J(t)(c_{n+1}+c_{n-1})-\mathscr{E}_0\cos(\omega t) x_n
c_n,
\end{eqnarray}
where the dot denotes the derivative with respect to time. To solve
Eq. (2), we make the function transformation $c_n(t)=A_n(t)
\exp(i\mathscr{E}_0\omega^{-1}x_n\sin\omega t)$ which leads Eq.(2)
to the form
\begin{eqnarray}
i\dot{A}_n(t)=J(t)(A_{n+1}e^{i \bigtriangleup_n\sin\omega
t}+A_{n-1}e^{-i \bigtriangleup_{n-1}\ \sin\omega t}).
\end{eqnarray}
In this equation, we have defined
$\bigtriangleup_n=\frac{\mathscr{E}_0}{\omega}(x_{n+1}-x_n)$ such
that there are the values $\bigtriangleup_{n}=\frac{\mathscr{E}_0}{\omega}a,\ \bigtriangleup_{n-1}=\frac{\mathscr{E}_0}{\omega}b$ for even $n$, and  $\bigtriangleup_{n}=\frac{\mathscr{E}_0}{\omega}b,\ \bigtriangleup_{n-1}=\frac{\mathscr{E}_0}{\omega}a$ for odd $n$.

We focus our attention on the situation of high-frequency regime
with $\omega\gg 1$. The selective CDT has been illustrated
analytically and numerically under this limit \cite{Creffield2}. We
shall give a general analytical solution of the system, which reveals the phase-controlled CDT and directed transport. Note that in Eq. (3),
$A_n(t)$ may be treated as a set of slowly varying functions
of time, and the coupling function $F(t,\bigtriangleup_n)=J(t)e^{i\bigtriangleup_n\sin \omega t}=\{J_0+\frac {1}{ 2} \delta J[e^{i (m\omega t-\phi)}+ e^{-i (m\omega t-\phi)}]\}\sum_{n'}\mathcal {J}_{n'}(\bigtriangleup_n)e^{i n' \omega t}$ is a rapidly oscillating function. Thus the function $F(t,\bigtriangleup_n)$ in Eq. (3) can be replaced by its
time-average
\begin{eqnarray}
&& \overline {F}(m,\phi,\bigtriangleup_n) \nonumber \\
&=& J_0\mathcal {J}_{0}(\bigtriangleup_n)
+ \frac 1 2 \delta J [ e^{i\phi}+(-1)^m e^{-i\phi}] \mathcal {J}_{m}(\bigtriangleup_n)\nonumber \\
&=&J_0\mathcal {J}_{0}(\bigtriangleup_n)+\left\{\begin{array}{ll} \delta J \cos \phi \mathcal {J}_{m}(\bigtriangleup_n)
\\  \\
i \delta J \sin \phi \mathcal {J}_{m}(\bigtriangleup_n)
 \end{array}\right. \ \  \begin{array}{ll} \textrm{for even}\ m,
\\  \\  \textrm{for odd}\ m, \end{array}
\end{eqnarray}
which just is the effective tunneling rate with $\mathcal {J}_{m}(\bigtriangleup_n)=(-1)^m \mathcal {J}_{-m}(\bigtriangleup_n)=(-1)^m \mathcal {J}_{m}(-\bigtriangleup_n)$ be the $m$th Bessel
function of the first kind \cite{khai}. Similarly, the time-average of $F(t,-\bigtriangleup_{n-1})=J(t)e^{-i\bigtriangleup_{n-1}\sin \omega t}$ in Eq. (3) reads $\overline {F}(m,\phi,-\bigtriangleup_{n-1})$, which is evaluated from Eq. (4) by using $-\bigtriangleup_{n-1}$ instead of $\bigtriangleup_{n}$. For an even (odd) $n$, $\overline {F}(m,\phi,\bigtriangleup_n)$ and $\overline {F}(m,\phi,-\bigtriangleup_{n-1})$ are associated with the effective tunneling rates of the lattice separations $a\ (b)$ and
$b\ (a)$, respectively. Clearly, they may be real or complex, corresponding to the even or odd $m$. Given Eq. (4), Eq. (3) is transformed to
\begin{eqnarray}
i\dot{A}_n(t)=\overline {F}(m,\phi,\bigtriangleup_n) A_{n+1}+\overline {F}(m,\phi,-\bigtriangleup_{n-1})A_{n-1}.
\end{eqnarray}
In the case of infinite-site lattice,
we can construct the exact general solution of Eq. (5) by applying the discrete Fourier transformation $A(k,t)=\sum_n A_n(t)e^{-i n k}=A_e(k,t)+A_o(k,t)$ to transform Eq. (5) into the equations $i\dot{A}_e(k,t)=\overline{f}(k)A_o(k,t),\ i\dot{A}_o(k,t)=\overline{f}^*(k)A_e(k,t)$ with $A_e$ and $A_o$ being the sums of even terms and odd terms respectively in the Fourier series. It should be reminded in the calculations that $\bigtriangleup_{n}=\frac{\mathscr{E}_0}{\omega}a,\ \bigtriangleup_{n-1}=\frac{\mathscr{E}_0}{\omega}b$ for even $n$, and  $\bigtriangleup_{n}=\frac{\mathscr{E}_0}{\omega}b,\ \bigtriangleup_{n-1}=\frac{\mathscr{E}_0}{\omega}a$ for odd $n$. From the two first order equations of $A_e$ and $A_o$ we derive the second order equation $\ddot{A}(k,t)=-|\overline{f}(k)|^2 A(k,t)$ with the well-known general solution $A(k,t)=\alpha(k)
e^{i|\overline{f}(k)|t}+\beta(k) e^{-i|\overline{f}(k)|t}$. Inserting this solution into the inverse transformation $A_n(t)=\frac{1}{2\pi}\int_{-\pi}^{\pi} A(k,t)e^{i n k}dk$, we immediately obtain the general solution of Eq. (5) as \cite{Dunlap,khai}
\begin{eqnarray}
A_n(t)=\frac{1}{2\pi}\int_{-\pi}^{\pi}[\alpha(k)
e^{i|\overline{f}(k)|t}+\beta(k) e^{-i|\overline{f}(k)|t}]e^{ink}dk.
\end{eqnarray}
Here $\overline{f}(k)$ takes the form
\begin{eqnarray}
&& \overline{f}(k)=J_+ \cos k+i J_- \sin k,   \nonumber
\\  && J_{\pm} = \overline {F}(m,\phi,\bigtriangleup_n)\pm \overline {F}(m,\phi,-\bigtriangleup_{n-1}),
\end{eqnarray}
$|\overline{f}(k)|$ and $\overline{f}^*(k)$ are the corresponding modulus and complex conjugate, $\alpha(k)$; $\beta(k)$ are adjusted by the initial conditions. Without
loss of generality, let the initially occupied state be $|\psi (t_0=0)\rangle=|N \rangle$ with a fixed integer $N$, namely the initial conditions read $A_N(0)=1,\ A_{n\ne N}(0)=0$. Combining the conditions with the discrete Fourier transformation and general solution of $A(k,t)$ produces
$A(k,0)=\alpha(k)+\beta(k)=e^{-iNk},\ i\dot {A}(k,0)=-|\overline{f}(k)|[\alpha(k)-\beta(k)]=i\sum_{n'}\dot{A}_{n'}(0)e^{-in'k}=\overline {F}(m,\phi,\bigtriangleup_N)e^{-i(N+1)k}+\overline {F}(m,\phi,-\bigtriangleup_{N-1})e^{-i(N-1)k}$. The final equation is derived from the initial conditions and Eq. (5). Solving the two equations of $\alpha(k)$ and $\beta(k)$ yields
\begin{eqnarray}
\alpha(k)&=&\frac {1}{ 2|\overline{f}(k)|} \Big[|\overline{f}(k)|e^{-iNk}
- \overline {F}(m,\phi,\bigtriangleup_N)e^{-i(N+1)k} \nonumber
\\ &-& \overline {F}(m,\phi,-\bigtriangleup_{N-1})e^{-i(N-1)k} \Big],\nonumber\\
\beta(k)&=&\frac {1}{ 2|\overline{f}(k)|} \Big[|\overline{f}(k)|e^{-iNk} +\overline {F}(m,\phi,\bigtriangleup_N)e^{-i(N+1)k} \nonumber
\\ &+& \overline {F}(m,\phi,-\bigtriangleup_{N-1})e^{-i(N-1)k}\Big].
\end{eqnarray}
Given the general solution (6), we can investigate the general transport characterization for the different initial conditions $|\psi (t_0)\rangle$ and the different nonzero effective tunneling rates $\overline {F}(m,\phi,\bigtriangleup_n)$ and $\overline {F}(m,\phi,-\bigtriangleup_{n-1})$. In the general cases, the particle may be in the expanded states or localized states, depending on the system parameters.

\section{Unusual transport phenomena}

We are interested in the unusual transport phenomena such as the DL, CDT, instability and directed transport. It will be found that such unusual phenomena can be controlled under the initial condition $|\psi (t_0)\rangle$ and for some special parameter sets with different phases. The routes for implementing the phase-controlled transport are very different, compared to that of the previously considered case $\delta J=0$ with a constant tunneling rate $J_0$ \cite{Creffield2,khai}.

\emph{Phase-controlled CDT}. The CDT conditions mean the zero effective tunneling rates $\overline {F}(m,\phi,\bigtriangleup_n)=\overline {F}(m,\phi,-\bigtriangleup_{n-1})=0$ in Eq. (5), and $\overline{f}(k)=0$ in  Eq. (7). Substituting the latter into Eq. (6), the probability amplitude becomes a constant $A_n(t)=A_n(t_0)$ determined by the initial conditions for any $n$, which means the occurrence of CDT. Inserting the CDT conditions into Eq. (4), we get $J_0\mathcal {J}_{0}(\bigtriangleup_n)+\delta J \cos \phi_0\mathcal {J}_{m}(\bigtriangleup_n)=J_0\mathcal {J}_{0}(\bigtriangleup_{n-1})+\delta J \cos \phi_0\mathcal {J}_{m}(\bigtriangleup_{n-1})=0$ for an even $m$, and $ J_0\mathcal {J}_{0}(\bigtriangleup_n)+i\delta J \sin \phi_0\mathcal {J}_{m}(\bigtriangleup_n)=J_0\mathcal {J}_{0}(\bigtriangleup_{n-1})-i\delta J \sin \phi_0\mathcal {J}_{m}(\bigtriangleup_{n-1})=0$ for an odd $m$. Therefore, we can arrive at or deviate from the CDT conditions by fixing the parameters $m,\ J_0,\ \delta J,\ \bigtriangleup_{n},\ \bigtriangleup_{n-1}$ and adjusting the phase to arrive at or deviate from the phase $\phi_0$. In the general case, $\mathcal {J}_{0}(\bigtriangleup_{n})\ne \mathcal {J}_{0}(\bigtriangleup_{n-1})$, the above CDT conditions imply $-\delta J \cos \phi/J_0=\mathcal {J}_{0}(\bigtriangleup_n)/ \mathcal {J}_{m}(\bigtriangleup_n)=\mathcal {J}_{0}(\bigtriangleup_{n-1})/\mathcal {J}_{m}(\bigtriangleup_{n-1})$ for an even $m$. For example, applying the parameters $J_0=1,\ \delta J=0.8,\ m=2$ to the CDT conditions produces the required values $\bigtriangleup_{n}\approx 2.01717,\ \bigtriangleup_{n-1}\approx 5.37977$. Adopting these parameters, we plot the effective tunneling rates as functions of phase, as in Fig. 2. It is shown that the effective tunneling rates are tunable by varying the phase, and the CDT conditions $\overline {F}(m,\phi,\bigtriangleup_n)=\overline {F}(m,\phi,-\bigtriangleup_{n-1})=0$ are established at the phase $\phi=\phi_0\approx 2.4$.

As a simplest example, we can fix the lattice separations $a,\ b$ and tune the ratio $\mathscr{E}_0/\omega$ to obey $\mathcal {J}_{0}(\bigtriangleup_{n})=\mathcal {J}_{0}(\bigtriangleup_{n-1})=0$ with $\bigtriangleup_{n}=\frac{\mathscr{E}_0}{\omega}a\approx 2.4048,\ \bigtriangleup_{n-1}=\frac{\mathscr{E}_0}{\omega}b\approx 5.5201$, then Eq. (4) becomes
\begin{eqnarray}
\overline {F}(m,\phi,\bigtriangleup_n)=\left\{\begin{array}{ll} \delta J \cos \phi \mathcal {J}_{m}(\bigtriangleup_n)
\\  \\
i \delta J \sin \phi \mathcal {J}_{m}(\bigtriangleup_n)
 \end{array}\right. \ \  \begin{array}{ll} \textrm{for even}\ m,
\\  \\  \textrm{for odd}\ m, \end{array}
\end{eqnarray}
Thus we can achieve the CDT by varying value of the phase to $\phi_0=\pi/2$ for an even $m$ or to $\phi_0=0$ for an odd $m$. The phase modulations may be performed in a nonadiabatic \cite{Creffield2,Romero} or adiabatic  manner \cite{Qian,SLonghi}.

\begin{figure}
\begin{center}
\includegraphics[width=2.5in]{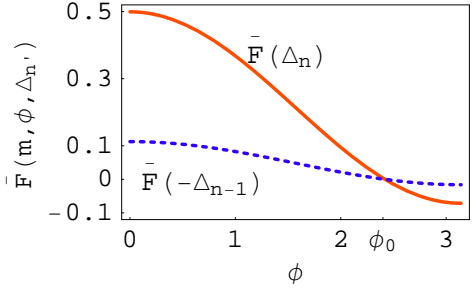}
\caption{The effective tunneling rates as functions of phase for the parameters $J_0=1,\ \delta J=0.8,\ m=2,\ \bigtriangleup_{n}=2.01717,\ \bigtriangleup_{n-1}=5.37977$. The solid and dashed curves describe $\overline {F}(m,\phi,\bigtriangleup_{n})$ and $\overline {F}(m,\phi,-\bigtriangleup_{n-1})$ respectively, which have the same zero point $\phi_0\approx 2.4$.}
\end{center}
\end{figure}

\emph{Phase-controlled DL}. The DL conditions mean one of the two effective tunneling rates vanishing.  When $\overline {F}(m,\phi,-\bigtriangleup_{n-1})=0$ and $|\psi (0)\rangle=|N \rangle$ are set, from Eqs. (7) and (8) we obtain the constant $|\overline{f}(k)|=|\overline {F}(m,\phi,\bigtriangleup_n)|$ and the periodic functions $\alpha (k)e^{ink}= \frac 1 2 [e^{i(n-N)k}-\overline {F}(m,\phi,\bigtriangleup_N)e^{i(n-N-1)k}/|\overline {F}(m,\phi,\bigtriangleup_n)|],\ \beta(k) e^{ink}= \frac 1 2[e^{i(n-N)k}+\overline {F}(m,\phi,\bigtriangleup_N)e^{i(n-N-1)k}/|\overline {F}(m,\phi,\bigtriangleup_n)|]$ of $k$. Inserting these into Eq. (6) produces the probability amplitudes
\begin{eqnarray}
A_{n\ne N, N+1}(t)=0,\ \ A_{N}(t)=\cos (\omega_1 t), \nonumber
\\ A_{N+1}(t)=-i\frac{ \overline {F}(m,\phi,\bigtriangleup_{N})}{|\overline {F}(m,\phi,\bigtriangleup_{N})|}\sin  (\omega_1 t).
\end{eqnarray}
They describe Rabi oscillation of the particle between the localized states $|N \rangle$ and $|N+1 \rangle$ with oscillating frequency $\omega_1 =|\overline {F}(m,\phi,\bigtriangleup_{N})|$. Similarly, taking $\overline {F}(m,\phi,\bigtriangleup_{n})=0$ leads to the probability amplitudes
\begin{eqnarray}
A_{n\ne N, N-1}(t)=0,\ \ A_{N}(t)=\cos  (\omega_2 t), \nonumber
\\ A_{N-1}(t)=-i\frac{ \overline {F}(m,\phi,-\bigtriangleup_{N-1})}{|\overline {F}(m,\phi,-\bigtriangleup_{N-1})|}\sin  (\omega_2 t),
\end{eqnarray}
which describe Rabi oscillation of the particle between the localized states $|N \rangle$ and $|N-1 \rangle$ with oscillating frequency $\omega_2 =|\overline {F}(m,\phi,-\bigtriangleup_{N-1})|$. Here the DL conditions and the different oscillating frequencies are modulated by the phase for a set of other parameters.

\emph{Phase-controlled instability}. Now we prove that the solutions of Eq. (5) are unstable under the condition
\begin{eqnarray}
\overline {F}(m,\phi_c,\bigtriangleup_{N})=-\overline {F}(m,\phi_c,-\bigtriangleup_{n-1})
\end{eqnarray}
for the phase $\phi=\phi_c$. In fact, when Eq. (12) is satisfied, Eq. (5) can be written as $d{A}_n(\tau)/d\tau = \frac 1 2 [A_{n-1}(\tau)-A_{n+1}(\tau)]$ with $\tau=2 i \overline {F}(m,\phi,\bigtriangleup_{N})t$, whose general solution is well-known as $A_n(\tau)=B_n \mathcal {J}_{n}(\tau)+D_n \mathcal {N}_{n}(\tau)$ with $\mathcal {J}_{n}(\tau)$ and $\mathcal {N}_{n}(\tau)$ being the Bessel and Neuman functions respectively, and $B_n,\ D_n$ the expansion coefficients determined by means of the initial conditions. For the complex variable $\tau$ with nonzero imaginary part, the asymptotic property $\lim_{\tau\to\infty}A_{n}(\tau)=\infty$ implies the instability of the solutions. Obviously such an instability can be controlled by tuning the phase to arrive at or deviate from $\phi_c$ in the condition (12), as shown in Fig. 3(a).

\emph{Phase-controlled directed transport}. For a set of given parameters $J_0,\ \delta J,\ \bigtriangleup_{n}$ and $\bigtriangleup_{n-1}$, we define two different phases $\phi_1$ and $\phi_2$ to obey
\begin{eqnarray}
&& \overline {F}(m,\phi_1,-\bigtriangleup_{n-1})\nonumber \\ & =& J_0\mathcal {J}_{0}(-\bigtriangleup_{n-1})+\delta J \cos \phi_1 \mathcal {J}_{m}(-\bigtriangleup_{n-1})=0, \nonumber
\\ && \overline {F}(m,\phi_2,\bigtriangleup_{n})\nonumber \\ & =&J_0\mathcal {J}_{0}(\bigtriangleup_n)+\delta J \cos \phi_2\mathcal {J}_{m}(\bigtriangleup_n)=0
\end{eqnarray}
for an even $m$, which result in the selective CDT between the localized states $|n \rangle$ and $|n-1 \rangle$ or between the localized states $|n \rangle$ and $|n+1 \rangle$, respectively. By nonadiabatically tuning the phase to alternately change between $\phi_1$ and $\phi_2$ just after the two different time intervals $T_1=\pi/\omega_1$ and $T_2=\pi/\omega_2$, the original tunneling rate becomes the continuous and piecewise analytic function $J(t,\phi)=J(t,\phi_1)$ for $t\in [n'(T_1+T_2),n'(T_1+T_2)+T_1]$, $J(t,\phi)=J(t,\phi_2)$ for $t\in [n'(T_1+T_2)+T_1,(n'+1)(T_1+T_2)]$ with $n'=0,1,2,...$. The instability condition (12) will be reached in each process changing phase from $\phi_i$ to $\phi_j$ for $i,j=1,2$.
\begin{figure}
\begin{center}
\includegraphics[width=2.5in]{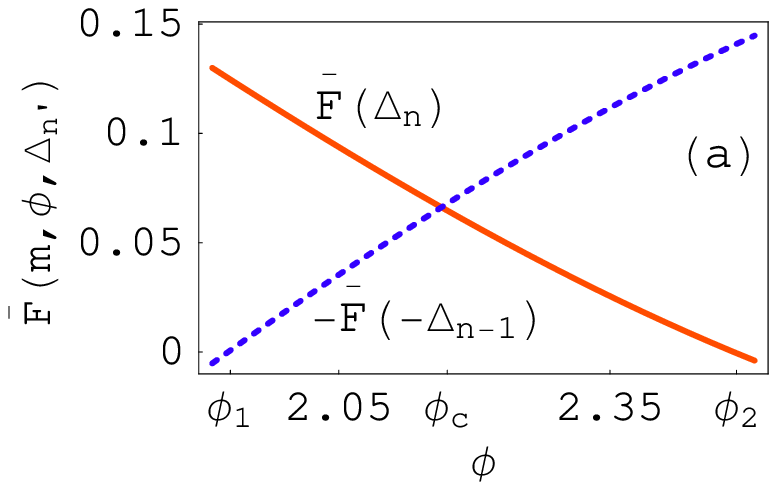}
\includegraphics[width=2.5in]{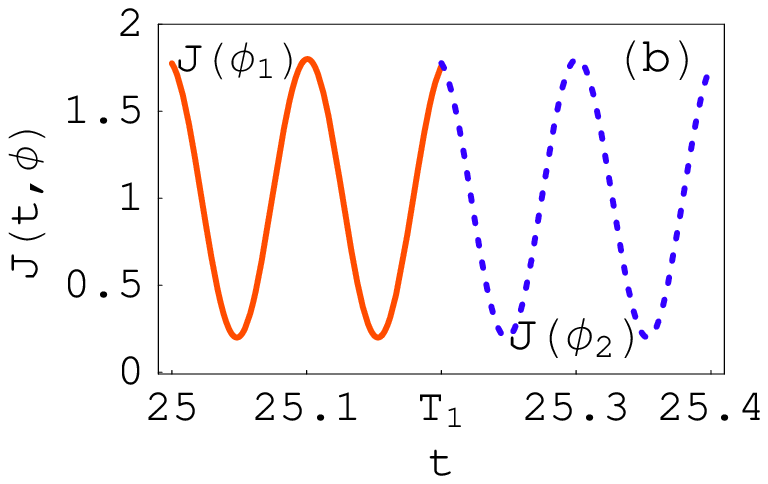}
\caption{The effective tunneling rates versus phase (a) and the time evolutions of the original tunneling rates (b) for the parameters $J_0=1,\ \delta J=0.8,\ m=2,\ \omega=30,\ \bigtriangleup_{n}=2,\ \bigtriangleup_{n-1}=2.2$. In (a), the solid curve describes $\overline {F}(m,\phi,\bigtriangleup_{n})$ with zero point $\phi_2\approx 2.49$  and the dashed curve labels $-\overline {F}(m,\phi,-\bigtriangleup_{n-1})$ with zero point $\phi_1\approx 1.93$ . The phase value $\phi_c \approx 2.17$ corresponds to the cross point of the two curves, where the instability condition (12) holds. In (b), the solid and dashed curves are associated with the original tunneling rates $J(t,\phi_1)$ and $J(t,\phi_2)$, respectively. At the time $t=T_1=\pi/\omega_1=25.2001$, the $J(t,\phi_1)$ is nonadiabatically changed to $J(t,\phi_2)$.}
\end{center}
\end{figure}
As an example, this is indicated by $\phi_c$ in Fig. 3(a) for the parameter set $J_0=1,\ \delta J=0.8,\ m=2,\ \omega=30,\ \bigtriangleup_{n}=2,\ \bigtriangleup_{n-1}=2.2$, where Eqs. (12) and (13) give $\phi_c \approx 2.17,\ \phi_1\approx 1.93$ and $ \phi_2\approx 2.49$, and the Rabi frequencies and half-periods of Eqs. (10) and (11) read $\omega_1=\overline {F}(m,\phi_1,\bigtriangleup_{n})\approx 0.124666,\ \omega_2=|\overline {F}(m,\phi_2,-\bigtriangleup_{n-1})|\approx 0.140933$ and $T_1\approx 25.2001,\ T_2\approx 22.2914$, respectively. The corresponding adjustment to the tunneling rate $J(t,\phi)=1+0.8 \cos(60 t+\phi)$ is exhibited in Fig. 3(b) for the time interval including $t=T_1$, which only transforms the phase of $J(t,\phi)$ from $\phi_1$ to $\phi_2$ and does not change its magnitude. Clearly, at the time $t=T_1+T_2$, the tunneling rate will change from  $J(t,\phi_2)$ to $J(t,\phi_1)$. By repeatedly using such operations, the initially stable Rabi oscillation is broken under the conditions (12), then the instability is suppressed by the conditions (13) with phase $\phi_1$ or $\phi_2$ such that the particle is forced to transit repeatedly between the two stable oscillation states $|\psi_n (t)\rangle=A_n|n \rangle+A_{n+1}|n+1 \rangle$ and $|\psi_{n'} (t)\rangle=A_{n'}|n' \rangle+A_{n'+1}|n'+1 \rangle$ with the amplitudes and frequencies of Eqs. (10) and (11) for $(n,n')=(N,N+1);(N+1,N+2);...$ that lead to the directed motion toward the right \cite{Creffield2,khai}. Because the phase-transformation does not change the magnitude of $J(t,\phi)$, it supplies a more convenient method to manipulate the directed transport, which is different from the previous magnitude-modulation of the coupling \cite{Romero}.

\section{Conclusions and discussions}

We have investigated the coherent control of single
particles held in the bipartite lattice with two different
separations $a$ and $b$ and driven by a combined modulation of two resonant
external fields with a phase difference between the bias and coupling. In the high-frequency regime and NNTB approximation, we derive an analytical general solution of the time-dependent Schr\"odinger equation, which
quantitatively describes the dependence of the tunneling dynamics on the phase difference between the modulation components. It is demonstrated that a new route of CDT or DL can be formed by tuning the phase to make two or one of the effective tunneling rates of the lattice separations $a$ and
$b$ vanishing. When the two effective tunneling rates are
adjusted to go through the values of the same magnitude and opposite
signs, the system loses its stability. The phase-controlled selective CDT enables the system to be stabilized and the
directed tunneling of the particle to be coherently manipulated. In
the process of control, the appropriate operation times are fixed by
the two tunneling half-periods.

Experimentally, the periodic modulation of tunneling rate
was realized through a variation of the lattice depth \cite{Chen}. The phase-adjustment of the periodic shaking can be implemented adiabatically or nonadiabatically, and may be more convenient compared to the usual amplitude- and frequency-modulations. The directed tunneling is related to the ratchetlike effect
of quantum particles, which can be tested with existing experimental setups on the periodically tilted and shaken optical lattices and could be well suited to simulating the similar optical systems \cite{Longhi2,Dreisow} and solid-state systems \cite{Villas,Longhi}, e.g. a periodically tilted and shaken chain of coupled quantum dots.

Particularly, in the NNTB approximation, the results can be extended to
controlling the directed motion of a many-particle system.
The possible experiment can begin by loading a Bose-Einstein condensate into the long lattice of wave-vector $k_L$, then one can tune the lattice depths to make the atomic sample in the Mott insulating state with a single atom per well \cite{RMa,Chen}, so the interparticle onsite interaction is neglectable. Further one can divide every well into a double well by ramping up the short lattice of wave-vector $2k_L$ and tilt the double-well train, that achieve the load of single atoms into the ``left" sides of tilted double wells \cite{Chen} with distances among particles being equal to the separation summation $a+b$. Thus we can apply the DL or CDT conditions to suppress pairing or bunching the particles, and make the couple between nearest-neighbor double-wells the neglectable so that the above-mentioned results, such as the phase-controlled DL, CDT and directed transport, hold for the many-particle system. In the DL conditions, every particle of the many-body system synchronously performs the Rabi oscillation in one of the double-wells  that generates a macromolecule-like in the bipartite lattice. While the directed transport of the many particles will form a stronger particle current, compared to the single particle case.

$\\ $ \bf Acknowledgment\ \rm
This work was supported by the NNSF of China under Grant Nos. 11204027, 11175064 and 11205021, the Construct Program of the National Key Discipline,
and the Hunan Provincial NSF (11JJ7001).

\end{document}